\documentclass[12pt]{article}
\usepackage{epsfig}
\def\be{\begin{equation}}
\def\ee{\end{equation}}
\def\bea{\begin{eqnarray}}
\def\eea{\end{eqnarray}}
\usepackage{graphicx}% Include figure files

\catcode`\@=11
\def\lsim{\mathrel{\mathpalette\@versim<}}
\def\gsim{\mathrel{\mathpalette\@versim>}}
\def\@versim#1#2{\vcenter{\offinterlineskip
\ialign{$\m@th#1\hfil##\hfil$\crcr#2\crcr\sim\crcr } }}
\catcode`\@=12
\usepackage{axodraw}

\catcode`@=12
\begin{document}
\thispagestyle{empty}
\begin{flushright}
UCRHEP-T488\\
March 2010\
\end{flushright}
\vspace{1.0in}
\begin{center}
{\LARGE \bf Unifiable Supersymmetric\\
Dark Left-Right Gauge Model\\}
\vspace{1.2in}
{\bf Ernest Ma\\}
\vspace{0.2in}
{\sl Department of Physics and Astronomy, University of California,\\
Riverside, California 92521, USA\\}
\end{center}
\vspace{1.2in}
\begin{abstract}\
The recently proposed dark left-right gauge model, with $Z'$ and $W_R^\pm$ 
bosons at the TeV scale, is shown to have a simple supersymmetric extension 
which is unifiable. Its one-loop gauge-coupling renormalization-group 
equations are shown to have identical solutions to those of the minimal 
supersymmetric standard model.  It also has a rich dark sector, with at 
least three stable particles.
\end{abstract}

\newpage
\baselineskip 24pt

\noindent \underline{\it Introduction}~:~ In the conventional left-right 
gauge extension of the Standard Model (SM) of particle interactions, 
the $SU(2)_R$ fermion doublet $(\nu,e)_R$ pairs up with the usual 
$SU(2)_L$ fermion doublet $(\nu,e)_L$ through a Higgs bidoublet, so that 
both the electron $e$ and the neutrino $\nu$ obtain Dirac masses. 
Remarkably, this situation is not compulsory.  It is in fact possible to 
have a symmetry such that $\nu_R$ is not the Dirac mass partner of $\nu_L$.  
It becomes another particle entirely, call it $n_R$, and the same symmetry 
makes it a dark-matter fermion (scotino).  This intriguing scenario 
has been elaborated in some recent papers~\cite{klm09,m09,adhm10,klm10}.  
The left-right structure itself was discussed already 23 years 
ago~\cite{m87,bhm87} in the context of superstring-inspired $E_6$ models. 
Called the Alternative Left-Right Model (ALRM), it has the important 
property of no tree-level flavor-changing neutral currents.  This makes it 
possible for the $SU(2)_R$ breaking scale to be as low as a TeV, allowing 
both its charged $W_R^\pm$ and $Z'$ gauge bosons to be observable at the 
large hadron collider (LHC).  However, its relevance to dark matter was not 
considered until one year ago~\cite{klm09}.

In this paper, the latest version~\cite{klm10} of the Dark Left-Right 
Model (DLRM) is shown to have a simple supersymmetric extension with 
gauge-coupling unification.  The resulting one-loop renormalization-group 
equations turn out to have solutions identical to those of the Minimal 
Supersymmetric Standard Model (MSSM), as well as two previously proposed 
left-right extensions~\cite{m95-1,m95-2}.  It also has a rich dark sector, 
with at least three stable particles~\cite{cmwy07}.

\noindent \underline{\it Model}~:~ Consider the gauge group $SU(3)_C \times
SU(2)_L \times SU(2)_R \times U(1)$.  Following Ref.~\cite{klm10}, a new 
global U(1) symmetry $S$ is imposed so that the spontaneous breaking of 
$SU(2)_R \times S$ will leave the combination $L = S + T_{3R}$ unbroken.  
Under $SU(3)_C \times SU(2)_L \times SU(2)_R \times U(1) \times S$, the 
superfields transform as shown in Table 1.  Because of supersymmetry, 
the Higgs sector is doubled, in analogy to the transition from the SM 
to MSSM.  Another set of Higgs doublet superfields $\eta$ and a new set of 
charged Higgs singlet superfields $\chi$ are added to obtain gauge-coupling 
unification~\cite{m95-1,m95-2}.

\begin{table}[htb]
\caption{Particle content of proposed model.}
\begin{center}
\begin{tabular}{|c|c|c|}
\hline
Superfield & $SU(3)_C \times SU(2)_L \times SU(2)_R \times U(1)$ & $S$ \\
\hline
$\psi = (\nu,e)$ & $(1,2,1,-1/2)$ & $1$ \\
$\psi^c = (e^c,n^c)$ & $(1,1,2,1/2)$ & $-3/2$ \\
$\nu^c$ & $(1,1,1,0)$ & $-1$ \\ 
$n$ & $(1,1,1,0)$ & $2$ \\ 
\hline
$Q = (u,d)$ & $(3,2,1,1/6)$ & $0$ \\
$Q^c = (h^c,u^c)$ & $(3^*,1,2,-1/6)$ & $1/2$ \\
$d^c$ & $(3^*,1,1,1/3)$ & $0$ \\
$h$ & $(3,1,1,-1/3)$ & $-1$ \\
\hline
$\Delta_1$ & $(1,2,2,0)$ & $1/2$ \\
$\Delta_2$ & $(1,2,2,0)$ & $-1/2$ \\
\hline
$\Phi_{L1}$ & $(1,2,1,-1/2)$ & $0$ \\
$\Phi_{L2}$ & $(1,2,1,1/2)$ & $0$ \\
$\Phi_{R1}$ & $(1,1,2,-1/2)$ & $-1/2$ \\
$\Phi_{R2}$ & $(1,1,2,1/2)$ & $1/2$ \\
\hline
$\eta_{L1}$ & $(1,2,1,-1/2)$ & $-2$ \\
$\eta_{L2}$ & $(1,2,1,1/2)$ & $2$ \\
$\eta_{R1}$ & $(1,1,2,-1/2)$ & $-3/2$ \\
$\eta_{R2}$ & $(1,1,2,1/2)$ & $3/2$ \\
\hline
$\chi_1$ & $(1,1,1,-1)$ & $-2$ \\
$\chi_2$ & $(1,1,1,1)$ & $2$ \\
\hline
\end{tabular}
\end{center}
\end{table}
The symmetry $S$ is used here to distinguish $\psi$, $\Phi_{L1}$, $\eta_{L1}$ 
from one another, as well as $\psi^c$, $\Phi_{R2}$, $\eta_{R2}$.  The bilinear 
terms allowed by $S$ are
\begin{equation}
\Delta_1 \Delta_2, ~~~ \Phi_{L1} \Phi_{L2}, ~~~ \Phi_{R1} \Phi_{R2}, ~~~ 
\eta_{L1} \eta_{L2}, ~~~ \eta_{R1} \eta_{R2}, ~~~ \chi_1 \chi_2.
\end{equation}
The trilinear terms are
\begin{eqnarray}
&& \psi \psi^c \Delta_1, ~~~ Q Q^c \Delta_2, ~~~ Q d^c \Phi_{L1}, ~~~ 
\psi \nu^c \Phi_{L2}, ~~~ n \psi^c \Phi_{R1}, ~~~ h Q^c \Phi_{R2}, \\
&& \Phi_{L1} \Phi_{R2} \Delta_2, ~~~ \Phi_{L2} \Phi_{R1} \Delta_1, ~~~ 
\eta_{L1} \eta_{R2} \Delta_1, ~~~ \eta_{L2} \eta_{R1} \Delta_2, \\ 
&& \Phi_{L1} \eta_{L1} \chi_2, ~~~ \Phi_{R1} \eta_{R1} \chi_2, ~~~ 
\Phi_{L2} \eta_{L2} \chi_1, ~~~ \Phi_{R2} \eta_{R2} \chi_1.
\end{eqnarray} 
Hence $m_e$ comes from the $I_{3L} = 1/2$ and $I_{3R} = -1/2$ component of 
$\Delta_1$ with $L=1/2-1/2=0$, $m_u$ from the $I_{3L} = -1/2$ and $I_{3R} = 
1/2$ component of $\Delta_2$ with $L=-1/2+1/2=0$, $m_d$ from $\phi^0_{L1}$, 
$m_\nu$ from $\phi^0_{L2}$, $m_n$ from $\phi^0_{R1}$, and $m_h$ from 
$\phi^0_{R2}$.  This structure guarantees the absence of tree-level 
flavor-changing neutral currents~\cite{gw77}.

\noindent \underline{\it Dark matter}~:~
As it stands, both the neutrino $\nu$ ($L=1$) and the scotino $n$ ($L=2$) 
are Dirac fermions, and lepton number $L$ is conserved.  If we now introduce 
a mass term $\nu^c \nu^c$ which breaks $L$ by two units, then $\nu$ gets a 
small Majorana mass through the canonical seesaw mechanism, as is usually 
assumed. As for $n$, it remains a Dirac fermion, being still protected by a 
global U(1) symmetry.  This can be understood by noting that with the 
addition of the $\nu^c \nu^c$ term, the same allowed bilinear and trilinear 
terms in Eqs.~(1) to (4) are obtained, if an odd matter parity $M$ is assumed 
for $\psi,\psi^c,\nu^c,n,Q,Q^c,d^c,h$ and the $S$ assignments of $\psi$, 
$\psi^c$, $\nu^c$ and $n$ are changed to $0$, $-1/2$, $0$, and $1$ 
respectively.  There are thus two conserved quantities: the usual $M$ (or its 
resulting $R$) parity and a global U(1) number $L' = S + T_{3R}$, with 
$L' = 0$ for the usual quarks and leptons and $L'=1$ for the scotino $n$.  
Because of their $S$ assignments, the $\eta$ and $\chi$ superfields appear 
always in pairs, so there is another parity $H$ which is conserved. Hence 
there are at least three stable particles.  Note that $\eta$ and $\chi$ 
communicate with the quarks and leptons only through $\Delta$ and $\Phi$, 
i.e. the so-called Higgs ``portals''.

The various superfields of this model under $L'$, $M$, and $H$ are listed 
in Table 2.  The usual $R$ parity is then defined as $R \equiv M H (-1)^{2j}$. 
A possible scenario for dark matter is to have the following three 
coexisting stable particles: the lightest neutralino ($L'=0$, $H=+$, $R=-$), 
the scotino $n$ ($L'=1$, $H=+$, $R=+$), and the exotic $\eta^0_{R2}$ fermion 
($L'=1$, $H=-$, $R=+$).  However, there may be additional stable particles 
due to kinematics.  For example, if the scalar counterpart of $n$ cannot 
decay into $n$ plus the lightest neutralino, then it will also be stable. 
There may even be several exotic stable $\eta$ and $\chi$ particles.  The 
dark sector may be far from just the one particle that is usually assumed, 
as in the MSSM.
\begin{table}[htb]
\caption{Superfields under $L'=S+T_{3R}$, $M$, and $H$.}
\begin{center}
\begin{tabular}{|c|c|c|c|}
\hline
$L'$ & $M$ & $H$ & Superfields\\
\hline
0 & $-$ & + & $u,d,\nu,e$ \\ 
0 & + & + & $g,\gamma,W_L^\pm,Z,Z'$ \\
0 & + & + & $\phi^0_{L1},\phi^-_{L1},\phi^+_{L2},\phi^0_{L2},\phi^0_{R1},
\phi^0_{R2}$ \\
0 & + & + & $\delta^0_{11},\delta^-_{11},\delta^+_{22},\delta^0_{22}$ \\
\hline
1 & $-$ & + & $n,h^c$ \\ 
$-1$ & $-$ & + & $n^c,h$ \\ 
1 & + & + & $W_R^+,\phi^+_{R2},\delta^+_{12},\delta^0_{12}$ \\ 
$-1$ & + & + & $W_R^-,\phi^-_{R1},\delta^0_{21},\delta^-_{21}$ \\ 
\hline
1 & + & $-$ & $\eta^0_{R2}$ \\ 
$-1$ & + & $-$ & $\eta^0_{R1}$ \\ 
2 & + & $-$ & $\eta^+_{L2},\eta^0_{L2},\eta^+_{R2},\chi^+_2$ \\ 
$-2$ & + & $-$ & $\eta^-_{L1},\eta^0_{L1},\eta^-_{R1},\chi^-_1$ \\ 
\hline
\end{tabular}
\end{center}
\end{table}

\noindent \underline{\it Gauge-coupling unification}~:~
The one-loop renormalization-group equations for the gauge couplings of 
$SU(3)_C \times SU(2)_L \times SU(2)_R \times U(1)_X$ are given by
\begin{equation}
{1 \over \alpha_i(M_1)} - {1 \over \alpha_i(M_2)} = {b_i \over 2 \pi} \ln 
{M_2 \over M_1},
\end{equation}
where $\alpha_i = g_i^2/4 \pi$ and the numbers $b_i$ are determined by the 
particle content of the model between $M_1$ and $M_2$.  In the SM with two 
Higgs scalar doublets, these are given by
\begin{eqnarray}
&SU(3)_C:& b_C = -11 + (4/3)N_f = -7, \\
&SU(2)_L:& b_L = -22/3 + (4/3)N_f + 2(1/6) = -3, \\ 
&U(1)_Y:& b_Y = (20/9)N_f + 2(1/6) = 7,
\end{eqnarray}
where $N_f=3$ is the number of quark and lepton families.  As such, the 
gauge couplings do not unify at a common mass scale, i.e. they do not 
satisfy the condition
\begin{equation}
\alpha_C(M_U) = \alpha_L(M_U) = (5/3)\alpha_Y(M_U) = \alpha_U.
\end{equation}
If the SM becomes the MSSM above $M_S$, the numbers $b_i$ change, i.e.
\begin{eqnarray}
&SU(3)_C:& b'_C = -11 + (2/3)(3) + (4/3+2/3)N_f = -3, \\
&SU(2)_L:& b'_L = -22/3 + (2/3)(2) + (4/3+2/3)N_f + (2/3+1/3)2(1/2) = 1, \\ 
&U(1)_Y:& (3/5)b'_Y = (4/3+2/3)N_f + (3/5)(2/3+1/3)(4)(1/4) = 33/5.
\end{eqnarray}
Therefore
\begin{eqnarray}
\ln {M_U \over M_Z} &=& {\pi \over 2} \left( {1 \over \alpha_L(M_Z)} - 
{1 \over \alpha_C(M_Z)} \right), \\ 
\ln {M_S \over M_Z} &=& {\pi \over 4} \left( {3 \over \alpha_Y(M_Z)} - 
{12 \over \alpha_L(M_Z)} + {7 \over \alpha_C(M_Z)} \right).
\end{eqnarray}
Now~\cite{pdg08}
\begin{eqnarray}
\alpha(M_Z)^{-1} &=& 127.953 \pm 0.049, \\ 
\sin^2 \theta_W(M_Z) &=& 0.23119 \pm 0.00014, \\ 
\alpha_L(M_Z) &=& \alpha(M_Z)/\sin^2 \theta_W(M_Z) = 0.03381, \\ 
\alpha_Y(M_Z) &=& \alpha(M_Z)/\cos^2 \theta_W(M_Z) = 0.01017, \\ 
\alpha_C(M_Z) &=& 0.1215 \pm 0.0017.
\end{eqnarray}
For $M_S > M_Z$, using Eq.~(14),
\begin{equation}
\alpha_C < {7 \alpha_L \alpha_Y \over 3 (4 \alpha_Y - \alpha_L)} = 0.1168,
\end{equation}
in disagreement with Eq.~(19).  However, this problem is usually fixed by 
going to two loops and spreading out the SUSY particle thresholds.

Consider now the dark left-right model.  At $M_R$, there is 
the boundary condition
\begin{equation}
{1 \over \alpha_Y(M_R)} = {1 \over \alpha_R(M_R)} + {1 \over \alpha_X(M_R)} 
= {1 \over \alpha_L(M_R)} + {1 \over \alpha_X(M_R)}.
\end{equation}
Above $M_R$, assuming the minimal supersymmetric content, 
without the exotic $\eta$ and $\chi$ superfields, 
\begin{eqnarray}
b_C &=& -11 + (2/3)(3) + (2+1)N_f = 0, \\
b_{L,R} &=& -22/3 + (2/3)(2) + (4/3+2/3)N_f + (2/3+1/3)(6)(1/2) 
= 3, \\ 
(3/2)b_X &=& (2+1)N_f + (1+1/2)(8)(1/4) = 12.
\end{eqnarray}
As such, again the gauge couplings do not unify, i.e. they do not satisfy
\begin{equation}
\alpha_C(M_U) = \alpha_L(M_U) = \alpha_R(M_U) = (2/3)\alpha_X(M_U) = \alpha_U.
\end{equation}
However, this may easily be changed with the addition of new 
particles~\cite{m05}.  With the $\eta$ superfields above $M_R$,
\begin{eqnarray}
b'_C &=& -11 + (2/3)(3) + (2+1)N_f = 0, \\
b'_{L,R} &=& -22/3 + (2/3)(2) + (4/3+2/3)N_f + (2/3+1/3)(8)(1/2) 
= 4, \\ 
(3/2)b'_X &=& (2+1)N_f + (1+1/2)(16)(1/4) = 15,
\end{eqnarray}
and the $\chi$ superfields above $M_X$,
\begin{equation}
(3/2)b''_X = (2+1)N_f + (1+1/2)(16)(1/4) + (1+1/2)(2)(1) = 18.
\end{equation}
The resulting solutions are
\begin{eqnarray}
\ln {M_U \over M_Z} &=& {\pi \over 2} \left( {1 \over \alpha_L(M_Z)} 
- {1 \over \alpha_C(M_Z)} \right), \\ 
\ln {M_R^7 \over M_X^3 M_Z^4} &=& \pi \left( {3 \over \alpha_Y(M_Z)} 
- {12 \over \alpha_L(M_Z)} + {7 \over \alpha_C(M_Z)} \right).
\end{eqnarray}
Note that Eqs.~(30) and (31) are identical to Eqs.~(13) and (14) of the MSSM 
respectively if we set $M_X=M_R=M_S$.  Thus this model is no worse 
than the MSSM for gauge-coupling unification, and two-loop equations  
and particle thresholds may be invoked to fix it.  The same one-loop 
solutions are obtained in two other previously proposed supersymmetric 
left-right models~\cite{m95-1,m95-2}.  If the one-loop equations (30) and 
(31) are taken at face value, $M_U = 3.33 \times 10^{16}$ GeV and 
$M_R^{7/4} M_X^{-3/4} = 14.7$ GeV.  Take for example $M_R = 500$ GeV, then 
$M_X = 55.2$ TeV.  The additional singlets are thus much heavier, but 
since they do not affect either $SU(2)_R \times U(1)_X$ breaking or 
supersymmetry breaking, this is an acceptable scenario.
In terms of $SO(10)$ multiplets, $Q,Q^c,\psi,\psi^c$ belong to \underline{16}; 
$h,d^c$ and $\Delta$ to \underline{10}; $\Phi$ and $\eta$ to \underline{16} + 
\underline{16}$^*$; and $\chi$ to \underline{120} + \underline{120}$^*$. 
The chosen set of superfields is free of anomalies.

\noindent \underline{\it Conclusion}~:~
The dark left-right model, where the $SU(2)_R$ fermion doublet $(n,e)_R$ 
contains the dark-matter fermion (scotino) $n$ which is distinguished from 
the usual lepton $e$ by an U(1) global symmetry, is extended to include 
supersymmetry.  New superfields are added, resulting in one-loop 
renormalization-group equations for the gauge couplings with solutions 
identical to those of the MSSM.  The dark sector is automatically extended 
to include at least three stable particles.

\noindent \underline{\it Acknowledgement}~:~
This work was supported in part by the U.~S.~Department of Energy Grant 
No. DE-FG03-94ER40837.

\newpage
\bibliographystyle{unsrt}

\end{document}